**Fröhlich-type Polarons in Isotopically Enriched Hexagonal Boron Nitride**

Ioannis Chatzakis[1*], Timur Abdilov[1,2], Elliot Walker[3], Jaime Freitas[4], Song Liu[5], James H. Edgar[5]

[1]Department of Physics and Astronomy, Texas Tech University, Lubbock, TX 79409

[2]Department of Electrical Engineering, Texas Tech University, Lubbock, TX 79409

[3]Department of Mathematics, Texas Tech University, Lubbock, TX 79409

[4]U.S. Naval Research Laboratory, Washington, DC 20375

[5]Department of Chemical Engineering, Kansas State University, Manhattan, KS 66506

**Abstract**

Exciton-phonon interactions play a central role in defining the optical response of hexagonal boron nitride (hBN), yet their quantitative determination has remained incomplete. Here, we reveal the Fröhlich-type exciton-phonon coupling in boron-10-enriched hBN using low-temperature cathodoluminescence. We resolve the indirect exciton $5.95 \pm 0.02$ eV together with its longitudinal optical (LO) phonon replica detuned by $184 \pm 56\ meV$, enabling the extraction of a Fröhlich coupling constant $\alpha = 0.159$ and a larger exciton binding energy of $161\ meV$, larger than previously reported values for natural-abundance hBN, which is attributed to isotope enrichment. The inferred polaron radius exceeds the lattice constant, indicating large-polaron behavior. We deduced an exciton scattering time $\sim 97$ fs corresponding to a homogeneous linewidth of $\sim 6.76$ meV. We further obtain a polaron binding energy of magnitude $\sim 48$ meV and an effective mass of $1.045\, m_0$. These results provide a direct quantitative characterization of exciton–phonon coupling in isotopically engineered hBN and establish a foundation for tailoring its phonon-polaritonic and quantum-optical properties.

* Contact author: ioannis.chatzakis@ttu.edu

Hexagonal boron nitride (hBN) is a wide-bandgap (~6 eV) two-dimensional semiconductor with high thermal conductivity, excellent chemical stability, and strong optical anisotropy.[1,2] Its optical and electronic properties are strongly influenced by electron–phonon–exciton interactions[3,4,5] isotopic composition, and lattice defects, making it a versatile platform for diverse applications[6]. Defect engineering has enabled spin-dependent photoluminescence and optically detected magnetic resonance (ODMR)[7–9], analogous to nitrogen-vacancy centers in diamond[10], but accessible through scalable chemical vapor deposition (CVD) growth and irradiation methods. Beyond quantum sensing, hBN supports efficient deep-UV emission[11], hosts hyperbolic phonon–polaritons with deep subwavelength confinement[12–16], and enables the engineering of isotopically enriched $^{10}$B layers for compact solid-state neutron detectors[17]. When integrated with graphene and related 2D materials, hBN further enables the emergence of exotic physical phenomena[18]**.** These multifunctional properties position hBN as a key material bridging quantum technologies, nanophotonics, optoelectronics, and nuclear security[19,20].

A central challenge, however, remains in fully understanding how excitonic and phononic interactions govern its optical and quantum behavior, as this limited knowledge continues to constrain progress in device development. In particular, the role of longitudinal optical (LO) phonons[21] in exciton scattering and linewidth broadening remains a significant obstacle for engineering defect emission, enhancing UV luminescence, and improving spin coherence. We report the first direct experimental investigation of exciton–phonon interactions in hBN, demonstrating Fröhlich coupling[5,22–28,29] to LO phonons in quantitative agreement with theoretical predictions. By analyzing cathodoluminescence (CL) spectra in the 5.3–6.1 eV range, we identify excitonic peaks at cryogenic temperatures, extract the dimensionless Fröhlich constant, and determine the polaron radius-showing that excitons in hBN isotopically enriched [30,31,] with $^{10}$B form large polarons [32]. Isotopic enrichment reduces mass-disorder broadening and improves the resolution of phonon-assisted excitonic features. Additionally, we calculate the exciton scattering rate using the material's parameters adopted from the literature and find it in relatively good agreement with the experimentally obtained linewidths.

This systematic correlation between experiment and polaron theory provides quantitative insight into the strength of exciton–phonon coupling[23,25,33,34] in hBN. Such understanding is essential for the community to control linewidths and emission stability of excitons and defect centers for quantum sensing, optimize phonon-assisted recombination pathways for deep-UV photonic devices, and improve modeling of spin–lattice interactions for ODMR-based quantum technologies. By clarifying the microscopic origin of exciton–phonon scattering in hBN, this research helps overcome a major barrier to using hBN in scalable quantum and photonic applications.

Monoisotopic h$^{10}$BN crystals were synthesized using isotopically enriched boron-10 (99.22 at%) and a precursor solvent mixture consisting of 50 wt% Ni and 50 wt% Cr powders. The mixture was loaded into an alumina crucible, the furnace chamber evacuated and backfilled with $N_2$ and forming gas (5% hydrogen in balance argon) to a pressure of ~ 850 Torr. During crystal growth, nitrogen and forming gas were continuously supplied at flow rates of 125 sccm and 25 sccm, respectively. The system was heated to 1550 $^{0}$C for a dwell time of 24 h. The hBN crystals were formed by cooling at a rate of 1$^{0}$C/hour to 1500 $^{0}$C, then quenching the process to room temperature. Forming gas was employed to suppress oxygen and carbon incorporation, the dominant impurities in hBN crystals. Raman scattering measurements verified the high crystalline quality of the sample. The AFM edge profile indicates a thickness of $\sim 4\ \mu m$. The sample was exfoliated from a single crystal using standard exfoliation processes and transferred onto a sapphire substrate.

Isotopic enrichment enables controlled tuning of the LO-phonon energy via the $M^{-1/2}$ mass dependence while leaving the electronic structure largely unchanged, thereby providing a stringent platform for quantitative evaluation of Fröhlich exciton–phonon coupling. Raman spectra for both natural-abundance and $^{10}$B-enriched hBN are shown in Figs. 1-2 were acquired in our laboratory under identical experimental conditions. All cathodoluminescence spectra discussed in this work were measured on the isotopically enriched $^{10}$B hBN samples.

CL spectra with samples mounted on a cryogenic cold-finger inserted in an ultra-high vacuum chamber, which allows variable temperature measurements within the range $4 - 330\ K$. The film emissions were excited with an electron beam current of 3 μA and a beam accelerating voltage of 3 kV. The CL emission, collected with a combination of f-number-matching mirrors, was analyzed with an 85 cm Spex spectrometer. Experimental details are included in the Supplemental Material. The penetration depth $R$ of electrons in hBN is calculated using the diffusion model developed by Kanaya *et. al.*,[35] which considers electron scattering as a spherical diffusion process centered at the depth of maximum energy dissipation. This provides a simple way to relate backscattering and energy-loss phenomena to the geometry of electron transport in a material. The model is described by the relation.

$R = C \cdot V^{5/3}(1 + 0.978 \times 10^{-6} \cdot E)^{5/3}/(1 + 1.957 \times 10^{-6} \cdot E)^{5/3})^{4/8}$ , where $C = 2.76 \times 10^{-11} AE^{5/3}/Z^{8/9}\rho$, where $A = 24.02\ gr/cm^3$ (atomic weight) , $Z = 12$ (atomic number $B_{atom.numb.} + N_{atom.numb.}$), $\rho = 2.1\ gr/cm^3$ (mass density). For the acceleration voltage $V = 3\ KV$ we found a penetration depth of $R \sim 200nm$.

At high excitation powers, the luminescence efficiency of hBN saturates and drops significantly at shallow excitation depths. This behavior is likely related to surface effects, though the underlying mechanisms are not yet fully understood. In conventional $sp^3$ semiconductors such as diamond, surface dangling bonds provide non-radiative recombination pathways that reduce efficiency. Although dangling bonds are not expected in $sp^2$-bonded hBN, surface contamination, structural defects, or Auger recombination[36] may cause quenching. At high penetration depths, the luminescence efficiency saturates at the internal quantum yield, which reflects the fraction of excitons undergoing radiative recombination in the bulk. While luminescence efficiency depends on crystal orientation, surface termination, and contamination, quantum yield is an intrinsic bulk parameter. In diamond, indirect excitons recombine only via phonon-assisted processes, leading to low quantum yield. In contrast, hBN reaches an internal quantum yield of $\sim 50\%$ [37]. This unusually high efficiency for an indirect semiconductor suggests that nonradiative channels are strongly suppressed by fast radiative recombination pathways (a few hundred ps[38]).

## Results and Discussion

Figure 1 shows the Raman spectra of natural hBN and $h^{10}BN$ under 488 nm irradiation. The peak positions, determined by Lorentzian fitting, and appearing at $1365.4\ cm^{-1}$ ($\sim 169\ meV$) and $1393.2\ cm^{-1}$ ($\sim 173\ meV$), are consistent with previous reports[16,30,39]. The approximately 28 $cm^{-1}$ blue shift in $h^{10}BN$ arises from its lighter nuclear mass. In contrast, the Raman linewidths differ significantly: $8.02\ cm^{-1}$ for hBN and $3.27\ cm^{-1}$ for $h^{10}BN$. The full width at half maximum (FWHM) is related to phonon lifetime through the energy–time uncertainty principle. The corresponding to scattering times are $0.66\ ps$ for hBN and $1.62\ ps$ for $h^{10}BN$. The phonon lifetime $\tau$ includes contributions from anharmonic decay processes[40] into lower-energy phonons, among others, provided that energy and momentum are conserved. Additionally, Fig. 2 presents a two-dimensional Raman intensity map demonstrating a uniform distribution of $^{10}B$ within the sample. The $\sim 2\ cm^{-1}$ shift in peak position arises from the use of two different spectrometers for the two measurements in Figs. 1 and 2. The narrow distribution observed in the corresponding Raman histogram quantitatively confirms the high isotopic and structural homogeneity required for reliable exciton-phonon analysis. Figure 3 depicts the atomic force microscopy image of the sample, which verifies the $\sim 4\ \mu m$ thickness and confirms the bulk nature of our sample. Before proceeding with a quantitative analysis, we first outline the low-temperature 4 K CL spectra shown in Fig. 4. The peak centered at $\sim 4.1\ eV$ is due to a donor-acceptor pair (DAP) [41] recombination transition involving the nitrogen vacancy ($V_N$) shallow donor, and the deep acceptor formed by carbon impurity occupying the nitrogen ($C_N$) or boron ($C_B$) site, where the latter is suggested by Weston *et al.*[42] More recently, Plo *et al.*[43] introduced a method for identifying point defects using isotope substitution, polytype control, and experiment–theory comparison, and applied it to resolve the long-standing nature of the 4 eV color center in h-BN. Our results agree with the previous measurements of Du *et al.* [44], who assigned the shallow trap states to nitrogen vacancies $V_N$ (see

Fig. 1 in the Supplemental Material). Similarly to the natural hBN, the CL spectrum of $^{10}$B-enriched hBN in the range of $\sim 5.30 - 5.65\ eV$ is dominated by the defect-related lines, labeled D series[45,46], with only a minor contribution from lines above $5.65\ eV$ labeled S series. Below $5.65\ eV$, sharp defect emissions are strongly localized near dislocations and grain boundaries appear, in agreement with prior work[4,45,47–49]. Other lines are assigned to vacancies or defects[50]. For example, the $5.2695\ eV$ line is attributed to the $D_6$ boron nitride divacancy, while the $5.46\ eV$ line ($D_{3,4}$) originates from an excitonic state bound to structural defects[45]. The relative intensities of these lines vary with sample quality. In the D-series energy region, some peaks are attributed to phonon replicas[51], as shown in Fig. 2 of Pierret *et al.*[45]. Similar features appeared in previous studies reporting cathodoluminescence spectra for natural boron in hBN[45,52,53,54].

Furthermore, the phonon-assisted replicas in the $^{10}$B-enriched and natural hBN samples show a clear ~20 meV energy separation, with the replica in the $^{10}$B-enriched sample shifted to lower energies. This shift is consistent with the anticipated hardening of the optical phonon modes resulting from the reduced mass of $^{10}$B. Since the phonon frequency scales as $\omega \propto 1/\sqrt{M}$, the relative change follows $\Delta\omega/\omega \approx -\frac{1}{2}(\Delta M/M) \approx 4\%$ increase when going from natural boron to nearly pure $^{10}$B. For the dominant optical phonons involved in the hBN phonon replicas ($\approx 184\ meV$), such a $4 - 5\%$ increase in phonon energy corresponds to an expected shift of approximately $\Delta E_{\text{phonon}} \approx 7$–$15$ meV. The $\sim 20\ meV$ shift lies at the upper end of the theoretical range, reflecting the role of exciton–phonon coupling and confirming the pronounced sensitivity of phonon-assisted excitonic emission to isotopic composition in hBN.

Additional features at $6.02\ eV$ and $6.07\ eV$ correspond to optical absorption assisted by Transverse acoustic (TA) and longitudinal acoustic (LA) phonons.

The optical and electronic properties of hBN are governed by exciton–phonon interactions, dominated by the long-range Fröhlich coupling to polar optical phonons. This interaction underpins the extraordinary deep-UV emission that distinguishes hBN from other wide-bandgap materials. In polar crystals such as hBN, LO phonons generate macroscopic electric fields through collective charge oscillations, which exert long-range Coulomb forces on the exciton's electron-hole pair. The resulting coupling "dresses" the exciton with a phonon cloud, forming a polaron with an enhanced effective mass. These Fröhlich interactions enable highly efficient phonon-assisted recombination across the indirect bandgap. This unusual efficiency is exactly why hBN, despite its indirect band gap, functions as an outstanding deep-UV emitter. The long-range polar coupling efficiently couples excitons to LO phonons at the Brillouin zone edge, accelerating phonon-assisted recombination to a timescale of hundreds of picoseconds. This rapid radiative decay channel outcompetes many non-radiative processes, resulting in an extraordinary internal quantum efficiency[37] of up to approximately 50%, an unusually high value for an indirect-gap material. Furthermore, the scattering of excitons with phonons contributes to the exciton's spectral linewidth, particularly at elevated temperatures. Here, we focus on the exciton-LO phonon interaction. The spectral range $\sim 5.7$ to $6.2\ eV$ is illustrated in Fig. 5.

The emission feature at $\sim 5.95\ eV$ is attributed to the indirect exciton (iX), where electron–hole recombination requires phonon assistance to conserve momentum between inequivalent valleys. The lower-energy line at $\sim 5.77\ eV$ represents its LO-phonon replica, arising from coupling between the exciton and LO phonons via the Fröhlich interaction. A schematic in Fig. 6 illustrates the indirect transition, phonon wavevector, and phonon-assisted recombination process.

The lowest-energy indirect exciton (iX) involves an electron at the $\boldsymbol{M}$ point and a hole at the $\boldsymbol{K}$ point, giving the exciton a large center-of-mass momentum ($\boldsymbol{K_{exc}} \approx \boldsymbol{M} - \boldsymbol{K}$)[55]. Radiative recombination of this exciton requires momentum conservation via the simultaneous emission of a phonon with $\boldsymbol{q_{phonon}}$ ($\boldsymbol{K_{exc}} \approx \boldsymbol{q_{phonon}}$). The energy separation $\Delta E = E_{\text{exc}} - E_{\text{replica}}$ we determined from the constrained Voigt fits. Using the extracted peak positions $E_{\text{exc}} = 5.95 \pm 0.02\ meV$ and $E_{\text{replica}} = 5.77 \pm 0.05\ meV$ , we obtain $\Delta E = 184 \pm 56\ meV$ (see Fig.6), which falls within the literature-reported range for the $E_{1u}$ LO phonon in hBN ($\approx 162 - 201\ meV$)[24,56,57,58]. Notably, the phonon participating in the indirect recombination is not confined to the **Γ** point but samples the $E_{1u}$ LO branch along the **Γ**-**K** direction, where the phonon energy spans this range. We note that moderate lineshape asymmetry and partial peak overlap can introduce small systematic shifts in the fitted peak positions; however, the extracted $\Delta E$ remains stable within the quoted uncertainty for reasonable variations of the fitting model. Within this uncertainty, the measured detuning is consistent with the LO-phonon energy of hBN and supports the assignment of the observed feature to Fröhlich-type exciton–phonon coupling. Although the Fröhlich interaction is strongest for long-wavelength ($\boldsymbol{q} \approx 0$) phonons, the $E_{1u}$ LO mode in hBN exhibits energies in the range $162 - 201\ meV$ along the **Γ**-**M** direction[59] , while along **M**-**K** it is nearly dispersionless with energy $\sim 162\ meV$. Consequently, the phonon-assisted emission observed at an energy shifted by approximately the LO-phonon energy ($\hbar\omega_{LO} = 184\ meV$). This polar optical mode is distinct from the non-polar in-plane $E_{2g}$ Raman-active phonon ($\approx\ 173\ meV$ for $^{10}$B), which exhibits negligible Fröhlich coupling and does not contribute appreciably to the observed luminescence replica. We emphasize that the Raman-active $E_{2g}$ mode probes the zone-center phonon, whereas the phonon-assisted CL process involves the finite-$q$ $E_{1u}$ LO branch along the Γ–K direction.

In Fröhlich's model, the strength of the electron-phonon interaction is quantified by the dimensionless coupling constant $\alpha_F$ [24,34]

$$\alpha_F = \frac{e^2}{\hbar}\frac{1}{4\pi\epsilon_0}\left(\frac{1}{\varepsilon_\infty} - \frac{1}{\varepsilon_0}\right)\left(\frac{m_e^*}{2\hbar\omega_{LO}}\right)^{1/2} \quad , \tag{1}$$

where $m_e^*$ is the effective mass of the electron without the electron-phonon interaction, $\epsilon_0$ is the vacuum permittivity, $\hbar\omega_{LO}$ *is the* LO phonon energy, and $\varepsilon_\infty, \varepsilon_0$ *are the* dielectric constants for high and low frequency limits respectively. Similarly, the coupling constant for a hole can be calculated from Eq. (1) by replacing the mass $m^*$of the electron with the mass $m_h^*$ of the hole without the electron-phonon interaction. Here, considering the exciton-phonon interaction, the use of the single-carrier effective mass $m^*$ is invalid. Therefore, it is replaced by the exciton reduced $\mu_{exc.}$ denoted by $\frac{1}{\mu_{exc.}} = \frac{1}{m_e^*} + \frac{1}{m_h^*} = 0.26m_0$ , where $m_0$ is the electron mass $9.019 \times 10^{-31}kg$ . Then from Eq. (1), we deduce a Fröhlich constant $\alpha_{F\ exc} = 0.263$ , which is lower by a factor of 1.4 than the Fröhlich constant of an electron $\alpha_{F\ e} = 0.379$ for $\varepsilon_\infty = 4.98$ , *and* $\varepsilon_0 = 6.93$. For comparison, we also calculated the Fröhlich constant for the case in which, instead of using the exciton reduced mass, $\mu_{exc.}$ we used the sum of the electron and hole effective masses $M = (0.54 + 0.5)m_0 = 9.47 \times 10^{-31}kg$ , and obtained a value of 0.526.

The main parameter that determines the interaction of the electron (hole) with the optical phonon is the radius[24,26,28] given by

$$r_{e,h}^* = \left(\frac{\hbar}{2m_{e,h}\omega_{LO}}\right)^{1/2}, \tag{2}$$

where $m_{e,h}$ is the effective mass of the electron or the hole ($m_e = 0.54m_0$, $m_h = 0.50m_0$) [24]. Then, from Eq. (2), we infer $r_e^* = 6.29$Å for the electron and $r_h^* = 6.54$Å for the hole. Substituting $m_{e,h}$with the reduced mass $\mu_{exc.}$, we can rewrite Eq. (2) and define the exciton's polarization radius as $r_{pol} = (\hbar/2\mu_{exc.}\omega_{LO})^{1/2}$. We obtain a value of $r_{pol} = 9.09$Å, which is larger than the hBN lattice parameter, consistent with large polarons. The $r_{pol}$ determines the characteristic size of the polarization (polaron) radius, which is the spatial extent over which the electric field is induced, and the polarization accompanying a charge carrier.

The dimensionless Fröhlich constant using the characteristics of the Wannier-Mott exciton $E_b = \mu e^4/2(4\pi\varepsilon_0\varepsilon_\infty)^2\hbar^2$ (hydrogenic type) can also be expressed in terms of the exciton's binding energy $E_b$, the LO phonon energy, and the dielectric factor $(1 - \varepsilon_\infty/\varepsilon_0)$ [22,24,26,60] as

$$\alpha_{F\,ex} = \left(\frac{E_b}{\hbar\omega_{LO}}\right)^{1/2}\left(1 - \frac{\varepsilon_\infty}{\varepsilon_0}\right) \qquad (3)$$

Inverting Equation (3) yields the exciton's binding energy. For the single carrier (here for electron) mass, this energy is $E_{b\,e} = 337\,meV$, which is close to the reported value of $E_b \approx 350\,meV$ [37]. However, using the reduced mass $\mu_{exc.}$ in the calculation of $\alpha_{F\,ex}(= 0.263)$ the exciton results in a binding energy of $E_{b\,exc} = 161 meV$ . Our calculated exciton binding energy is somewhat higher than previously reported values in the literature (~130 meV, by Cassabois *et al.*,[1] and $149\,meV$ by Watanabe *et al*[61].,), but still within the typical range for 3D hBN. The larger exciton binding energy in the $^{10}$B -enriched hBN sample is a consequence of isotope-induced modifications of the lattice vibrational spectrum and the resulting changes in dielectric screening. Replacing the heavier $^{11}$B isotope with the lighter $^{10}$B increases the optical phonon frequencies ($\omega \propto m^{-1/2}$), thereby altering the ionic contribution to the static dielectric constant. Within the Lyddane–Sachs–Teller relation $\omega_{LO}^2/\omega_{TO}^2 = \varepsilon_0/\varepsilon_\infty$, [62] the static dielectric screening scales inversely with the square of the transverse optical phonon frequency, such that an increase in phonon energy reduces the lattice polarizability. Because exciton binding energies increase as the effective dielectric screening decreases ($E_b \propto 1/\varepsilon_{\text{eff}}^2$), the reduced ionic screening in the enriched sample leads to a stronger

electron–hole Coulomb attraction and thus a larger binding energy. This trend is consistent with recent first-principles studies showing that the lattice (phonon) contribution to exciton screening diminishes when the phonon frequencies shift upward, thereby enhancing exciton binding,[63,64,65] Our experimental observation of a $\sim 30\ meV$ increase in binding energy for the $^{10}$B-enriched material is therefore fully compatible with the expected reduction in phonon-mediated dielectric screening. These effects provide a consistent explanation for the enhanced binding energy Furthermore, we employed Feynman's theory to estimate the polaron's binding energy (self-energy)[66] and polaron mass. His path-integral method yields: $E_{pol}(\alpha) \cong -\hbar\omega_{LO}(\alpha_{F\,ex} + 0.0123\alpha_{F\,ex}{}^{2})$ which corresponds to the binding energy magnitude $|E_{pol}| \cong 48 meV$. The resulting polaron mass is $m_{pol}/m_e = \left(1 + \frac{\alpha}{6} + 0.025a^2 + \cdots\right) = 1.045$.

Notice that, as the exciton-phonon interaction constant depends on the optical phonon energy $\hbar\omega_{LO}$, which in turn depends on the isotopic mass, the isotopic substitution provides a viable method for the experimental tuning of exciton-phonon interactions. The phonon-assisted replica exhibits a pronounced shift of $\sim 20\ meV$ in the $^{10}$B-enriched sample. This behavior is consistent with the expected hardening of the optical phonon modes when the boron mass is reduced. Since the phonon frequency scales as ( $\omega \propto 1/\sqrt{M}$ ), the relative change follows $\Delta\omega/\omega \approx -\Delta M/2M \approx +4\%$ when going from natural boron to nearly pure $^{10}$B. For the dominant optical phonons involved in the hBN phonon replicas ($\approx 184\ meV$ ), such a $4-5\%$ increase in phonon energy corresponds to an expected shift of approximately $\Delta E_{phon} \approx 7-15\ meV$. The experimentally observed $\sim 20\ meV$ displacement falls at the upper edge of this theoretical range, indicating the exciton-phonon coupling and confirming the pronounced sensitivity of phonon-assisted excitonic emission to isotopic composition in hBN.

Moreover, we examine the effect of the Fröhlich interactions on the scattering rate of excitons by phonons. We begin from the Fermi-Golden rule[5] (Eq. 4) for LO-phonon emission by an exciton

$$\Gamma_{LO} = \frac{2\pi}{\hbar}\int \frac{d^3q}{(2\pi)^3}|g(q)|^2[N_{LO}+1]\delta\left(\frac{\hbar^2q^2}{2M} - \hbar\omega_{LO}\right), \qquad (4)$$

where $N_{LO}(T) = (e^{\hbar\omega_{LO}/k_BT} - 1)^{-1}$ is the Bose factor, and $|g(q)|^2 = \frac{e^2\hbar\omega_{LO}}{2\epsilon_0}\left(\frac{1}{\varepsilon_\infty} - \frac{1}{\varepsilon_0}\right)\frac{|F_{1s}(q)|^2}{q^2}$ is the Fröhlich matrix element including the 1s exciton form factor $|F_{1s}(q)|^2$, which removes the long-wavelength singularity at $q = 0$. The exact expression for the form factor is $F_{1s}(q) = \frac{1}{\left(1+\frac{r_{ex}^2q^2}{4}\right)^2}$, with $r_{ex}$ being the exciton radius, and takes the maximum value at $q = 0$. For the sum of the masses $M$, we obtain $q_0 = \sqrt{\frac{2M\omega_{LO}}{\hbar}} \approx 2.24 \times 10^9 m^{-1}\ (or\ 2.24\ nm^{-1})$. For $q = q_0\ and\ r_{ex} = 9.09$Å, the form factor $|F_{1s}(q_0)|$ is equal to $0.251$. Using the delta function property, $\delta(f(q) = \frac{\delta(q-q_0)}{|f'(q_0)|}$, in Eq. (4), we obtain an analytical formula (see Supplemental Material for details) to calculate $\Gamma_{LO}$,

$$\Gamma_{LO} = \frac{e^2\hbar\omega_{LO}(N_{LO}+1)M|F_{1s}(q)|^2}{2\pi\epsilon_0\hbar^2q_0}\left(\frac{1}{\varepsilon_\infty} - \frac{1}{\varepsilon_0}\right) \qquad (5)$$

which we find to be $1.71 \times 10^{13}s^{-1}$ where at 6 K the $(N_{LO}+1) \approx 1$. From the uncertainty principle, this corresponds to a homogeneous linewidth of $\Gamma \sim \frac{\hbar}{\tau_{LO}} \approx 11.3\ meV$.

Below, we experimentally determine the Fröhlich coupling constant $\alpha$ and compare it with the previously reported calculated values. The Fröhlich coupling constant $\alpha$ was determined experimentally by quantitatively fitting the temperature dependence of the zero-phonon excitonic emission linewidth. The spectra were analyzed using a constrained Voigt model (Lorentzian convoluted with Gaussian) to account for homogeneous and inhomogeneous broadening. Specifically, the number of spectral components was fixed based on the observed peak structure and prior literature assignments, and all peaks were fitted simultaneously so that neighboring components mutually constrain each other's widths; the individual Gaussian and Lorentzian widths within each Voigt profile were otherwise free to vary in the final fit. Under these conditions, the peak energies are stable, yielding $E_1 = 5.77 \pm 0.05$ eV and $E_2 = 5.95 \pm 0.02\ eV$. Peak energies are reported with conservative uncertainties reflecting the sampling interval, calibration accuracy, and fit stability. Given the wavelength-scan step ($\sim 0.9$ nm, corresponding to $\sim 20$ to $25\ meV$ near $6\ eV$), the Gaussian and Lorentzian components remain partially correlated at our sampling density; accordingly, their separation should be regarded as model dependent, whereas the total linewidth and peak positions are robust. We note that the narrowest spectral feature (indirect exciton) approaches the effective resolution set by the sampling interval. The extracted Lorentzian component therefore provides an effective estimate of the phonon-limited homogeneous broadening rather than a fully resolved intrinsic linewidth. The Lorentzian component is used as an effective homogeneous contribution consistent with the standard linewidth model. Importantly, the most robust quantity in our analysis – the energy separation between the exciton and its phonon replica – remains insensitive to the fitting model and robust within experimental uncertainty, thereby supporting the phonon assignment and the Fröhlich-coupling analysis. Peak energies are reported with conservative uncertainties reflecting the sampling density and statistical noise, while the absolute energy scale is accurate within $\pm 2 - 3\ meV$ based on Hg(Ar) calibration. The Lorentzian half-width corresponds to the intrinsic (homogeneous) component $\mathrm{FWHM}_{\mathrm{hom}}$. The temperature dependence of the total linewidth is described by the standard model[67] $\Gamma(T) = \Gamma_{\mathrm{inh}} +$ $a_{\mathrm{ac}}T + \Gamma_{\mathrm{LO}}/[\exp(\hbar\omega_{\mathrm{LO}}/k_BT) - 1]$. Here, $\Gamma_{\mathrm{inh}}$ is the temperature-independent inhomogeneous contribution, $a_{\mathrm{ac}}$ is the acoustic-phonon coupling coefficient, and the third term is the linewidth due to LO phonon scattering. $\Gamma_{\mathrm{LO}}$ is the LO-phonon scattering rate and the exciton lifetime is $\tau = 1/\Gamma_{\mathrm{LO}}$. At low temperature $(T = 6K)$, $[\exp(\hbar\omega_{\mathrm{LO}}/k_BT) - 1]^{-1} \approx 1$. From the fit to the indirect exciton emission at 5.95 eV, we obtain $\mathrm{FWHM}_{\mathrm{effective}} = 6.76$ meV. Using this effective homogeneous contribution, we estimate the LO-phonon scattering rate as $\Gamma_{\mathrm{LO}} = \mathrm{FWHM}_{\mathrm{effective}}/\hbar$ yielding $\Gamma_{LO} = 1.06 \times 10^{13}\mathrm{s}^{-1}$ and corresponding to $\tau \approx 97fs$. The instrumental broadening $(0.2 - 0.3\ meV$ for a $10\ \mu m$ slit) is negligible compared with the measured linewidth. While the Voigt fit yields a Gaussian component ($\sim 20.7\ meV$) and Lorentzian component ($\sim 6.76\ meV$), the total linewidth and its temperature dependence provide the most robust quantities for evaluating Fröhlich coupling. The fit of the spectra yielded a total linewidth of $24.4\ meV$, consisting of a Gaussian component of $20.7\ meV$ and a Lorentzian component of $\sim 6.76\ meV$ (see Supplemental Material for details). The Lorentzian FWHM obtained, $6.76\ meV$, is less than the $11.3\ meV$ predicted from Eq. (4). This difference arises because the theoretical value (Eq. 4) represents the total LO-phonon scattering rate in an ideal, perfectly homogeneous crystal, while the experimentally extracted Lorentzian width reflects only the phonon-limited component after convolution with significant inhomogeneous broadening (Gaussian = $20.7\ meV$). Furthermore, dielectric screening, finite exciton localization, and strain-induced potential fluctuations can reduce the effective exciton–phonon interaction[68,69]. Considering these effects, the measured homogeneous linewidth is consistent with a partially screened Fröhlich coupling in the investigated material. Then the Fröhlich constant given by $\alpha_{F\ ex} = \Gamma_{\mathrm{LO}}/\eta(q_0)\hbar\omega_{LO}$ yields $\alpha_{F\ ex} = 0.159$ for $\eta(q_0) = 0.241$ (See details in Supplemental Material). For a single carrier (electron), $\eta(q_0) = 0.167$, yielding $\alpha_{F\ e} = 0.219$, which is lower than the results from Sio *et al.,*. [24] by a factor ~ 1.4. Table 1 summarizes the results for different masses. Overall, the calculated values are larger than the experimental ones by an average factor of 1.4.

**Table 1:** The Fröhlich constant $\alpha_F$, calculated using Eq. (1), for the electron and hole effective masses, for the exciton's reduced mass, and for the sum of electron and hole effective masses (mass of the center of momentum of the exciton). Experimentally, the $\alpha_F$ is determined by the equation $\alpha_F = \Gamma_{\mathrm{LO}}/\eta\, \hbar\omega$.

| Mass | $\alpha_F$ calculated | $\alpha_F = \Gamma_{\mathrm{LO}}/\eta\, \hbar\omega$ |
|---|---|---|
| Exciton's Reduced mass $\frac{1}{\mu_{exc.}}$ | 0.263 | 0.159 |
| Effective hole mass $m_h^*$ | 0.364 | 0.210 |
| Effective electron mass $m_e^*$ | 0.379 | 0.219 |
| Total exciton mass $M$ | 0.526 | 0.314 |

**Conclusion**

In conclusion, we have demonstrated that isotopic engineering provides an effective route to tune exciton–phonon interactions in hexagonal boron nitride. Using cathodoluminescence spectroscopy, we show that $^{10}$B enrichment produces a systematic redshift of the excitonic emission lines and quantifiably modifies the indirect exciton and its LO-phonon replica. From a combined experimental–theoretical analysis, we extract the Fröhlich coupling strength, polaronic parameters, and exciton binding energy, finding clear signatures of large-polaron formation. The lighter $^{10}B$ isotope increases the optical phonon frequencies and reduces the ionic component of dielectric screening, thereby enhancing the electron–hole Coulomb interaction and yielding a larger exciton binding energy. These results establish isotopic substitution as a powerful tool for controlling polaronic and excitonic properties in wide-bandgap van der Waals materials.

**Acknowledgments**

We thank G. Cassabois at Laboratoire Charles Coulomb, Université de Montpellier, CNRS and Institut Universitaire de France for helpful discussions regarding the LO-phonon dispersion in hBN. We also thank Connie Li at U.S. Naval Research Laboratory for helping us with the AFM measurements. Ioannis Chatzakis acknowledges financial support from Texas Tech University startup funding. Support for h-BN crystal growth was provided by the National Science Foundation Award Number 1538127.

**Authors contributions**

The project was led by I.C., who conceptualized the work, analyzed the data, and wrote the manuscript. J.F. conducted the cathodoluminescence measurements. S.L. and J.E. grew the hBN crystals. T.A. and E.W. contributed to writing the manuscript. All authors contributed to improving the manuscript.

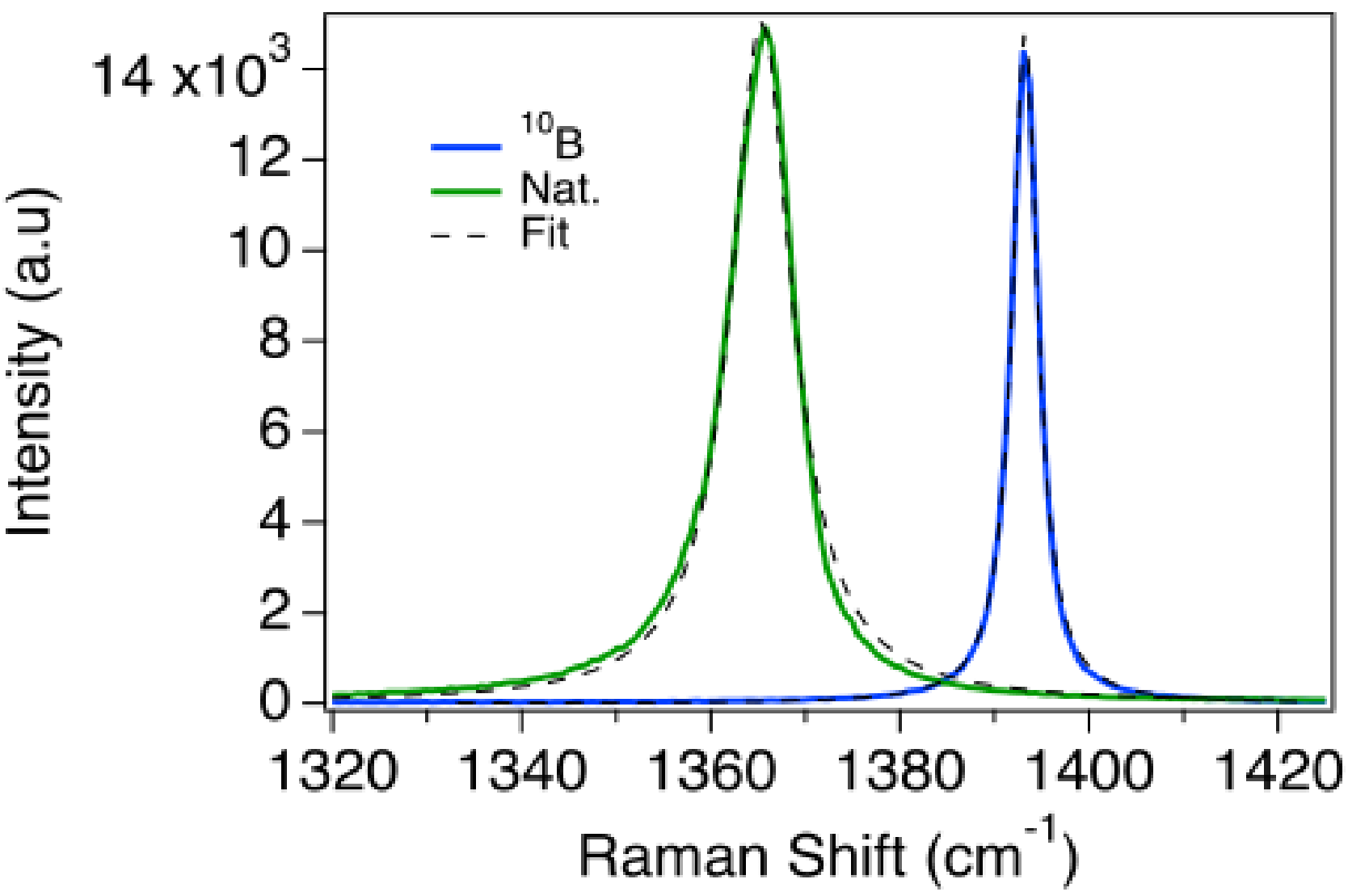

**Fig. 1** Raman spectrum for hexagonal boron nitride at 295 K. The green (natural) and blue ($^{10}$B isotope enriched) lines are the experimental data, and the dashed lines are the fits with Lorentzian function. The peak positions are at $1365\ cm^{-1}$ and $1393\ cm^{-1}$ respectively.

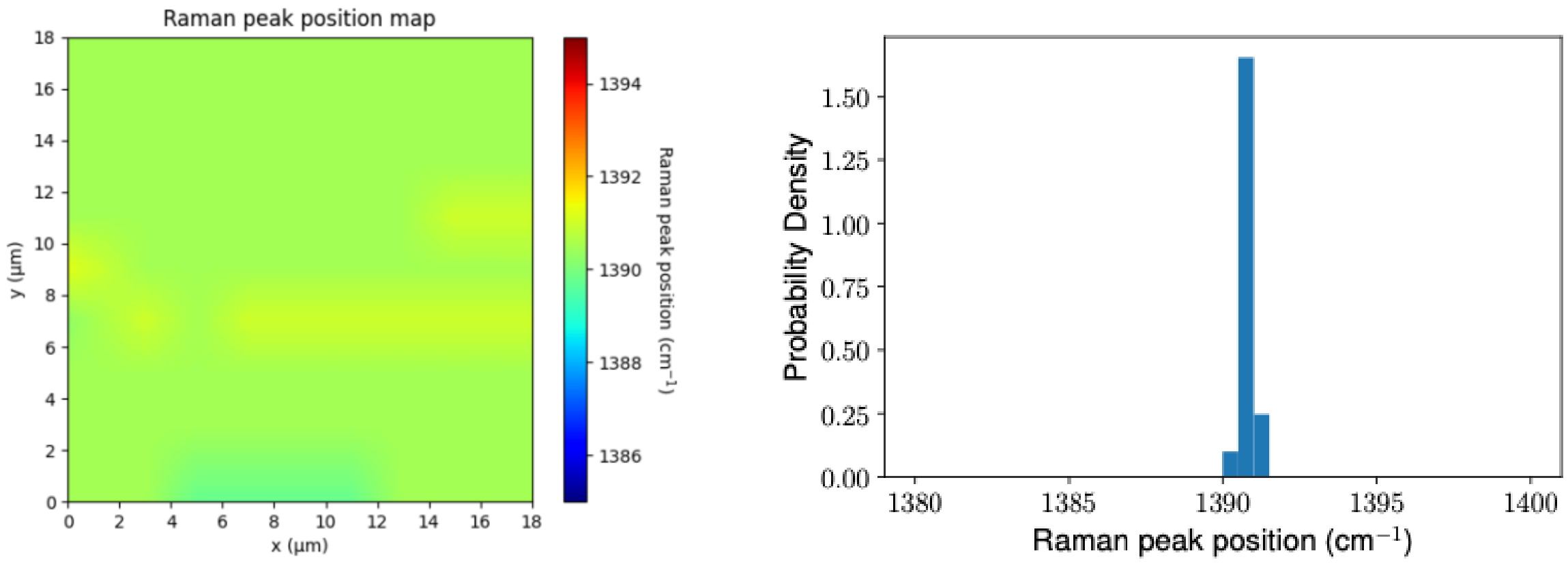

**Fig.2** (Left) Spatial Raman intensity map of the $E_{2g}$ mode of the $^{10}$B-enriched hBN crystal. The uniform contrast across the scanned region indicates high structural and isotopic homogeneity. (Right) Corresponding histogram of the Raman peak position extracted from the map, showing a narrow distribution centered at the characteristic phonon frequency. The small spread confirms the high spatial uniformity of the sample, which is essential for reliable quantitative analysis of exciton–phonon coupling.

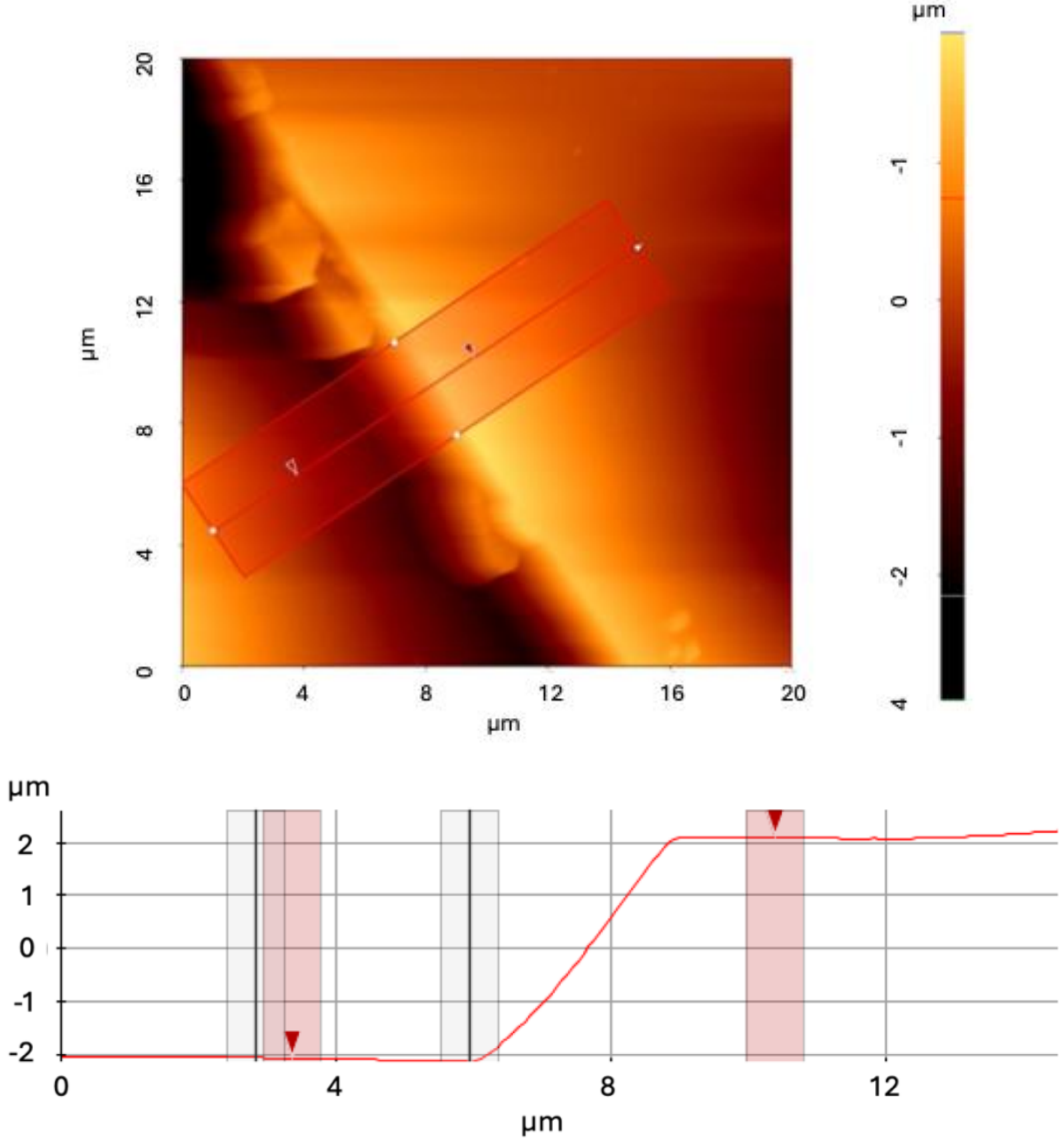

**Fig. 3** AFM image of the sample (top) and corresponding height profile (bottom) showing a thickness of ~4µm, confirming its three-dimensional character.

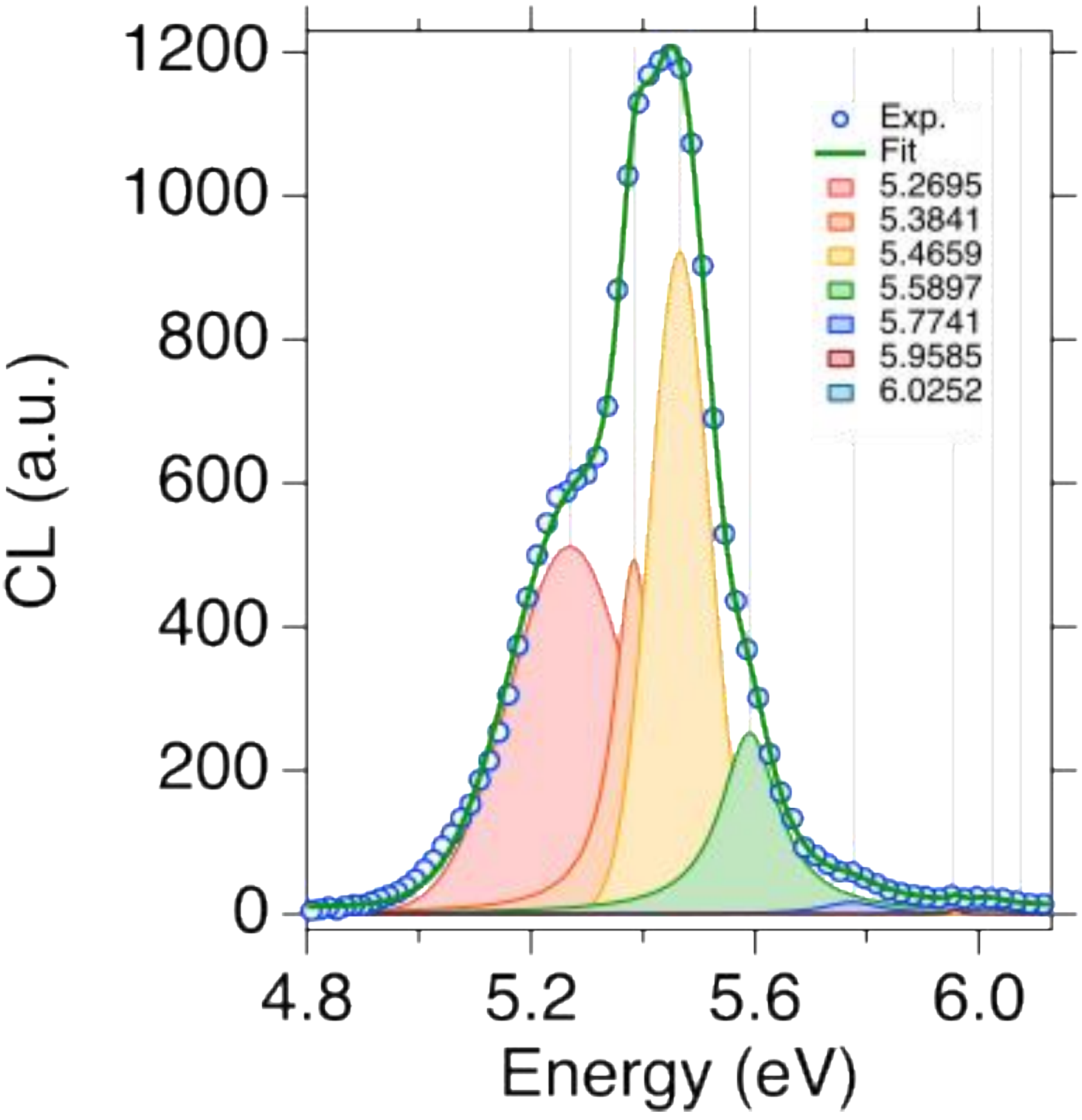

**Fig. 4** CL spectrum at 6 K. The large amplitude of the D series of lines in the ~ 5.2 to 5.77 eV range, compared to the peak amplitude above 5.8 eV, indicates the existence of structural defects. The positions of each emission band (in eV) are annotated with the corresponding colors.

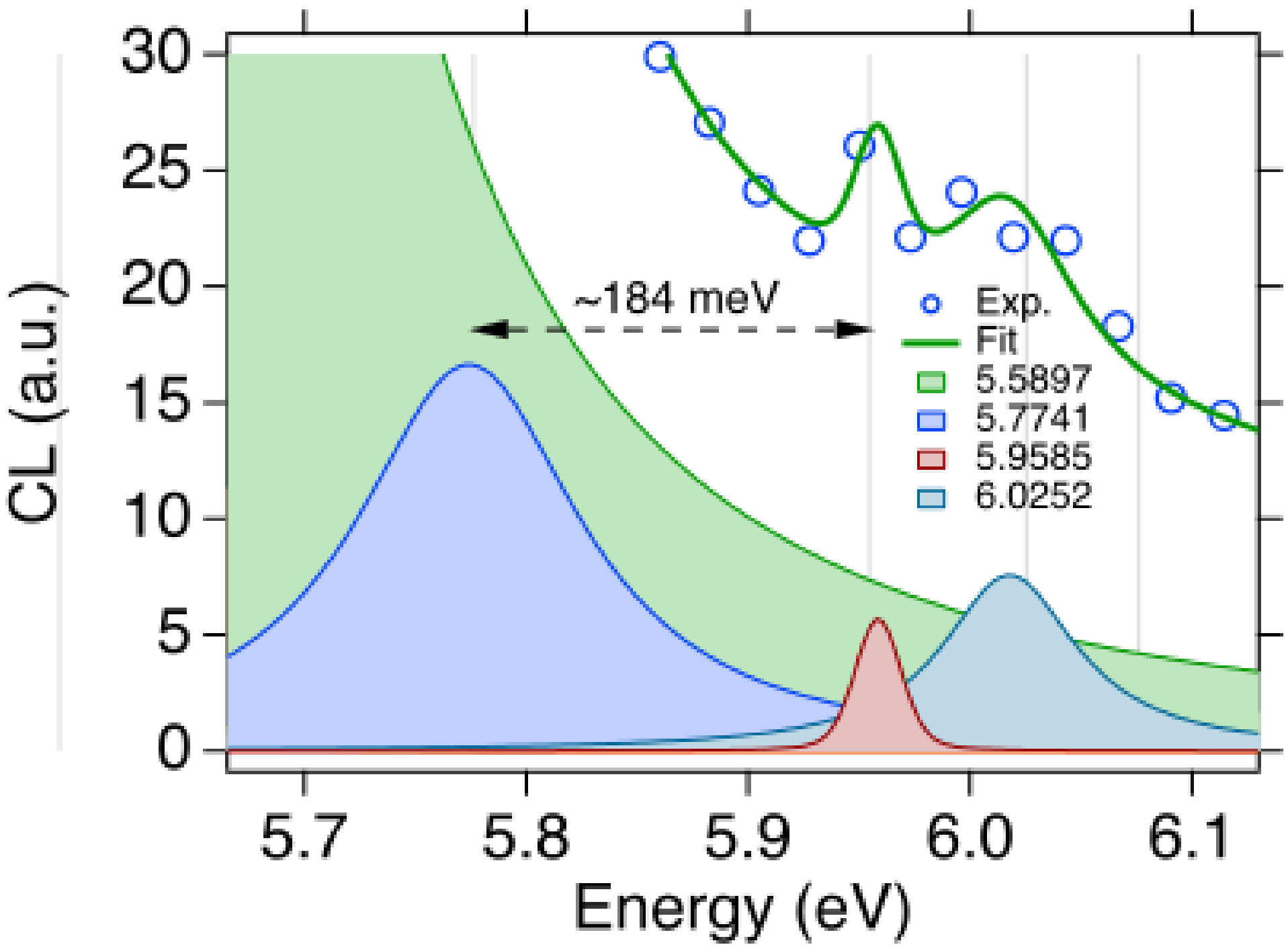

**Fig. 5** Expanded view of the high-energy region of Fig. 4. The green area highlights the transition located at ~5.56 eV. The brown color peak at ~ 5.95 eV corresponds to the indirect exciton, while its LO phonon mode replica, which appears at ~ 5.77 eV, is shown in blue. The transition at 6.02 eV is represented by the green-gray Lorentzian.

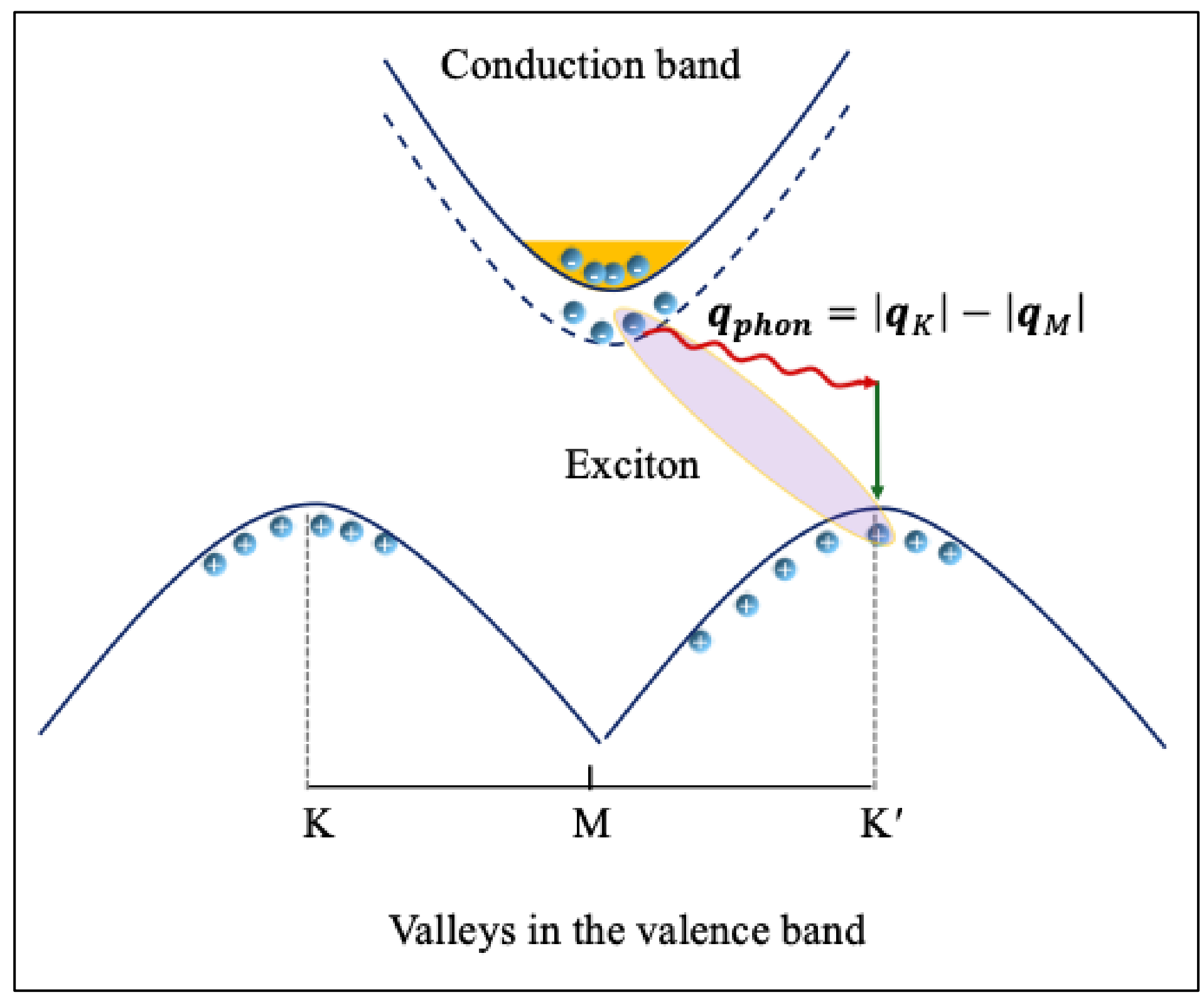

**Fig. 6** Schematic representation of the phonon-mediated indirect excitonic transition. The momentum required for the transition is provided by a phonon (red wavy line), whose scattering originates from the K–M or K′–M points of the Brillouin zone.